\documentclass[preprint,showpacs,preprintnumbers,amsmath,amssymb]{revtex4}
\usepackage{dcolumn}
\usepackage{bm}
\usepackage{CJK}
\usepackage{amsfonts}
\usepackage{psfrag}
\usepackage{wrapfig}
\usepackage{subfigure}
\usepackage{makeidx}
\usepackage{multirow}
\usepackage{epsf}
\usepackage{hyperref}
\usepackage{amsmath}

\usepackage{cases}

\begin{document}
\begin{CJK}{GBK}{song}

\newcommand{\eq}[1]{Eq.(\ref{#1})}
\newcommand{\ud}{\,\mathrm{d}\,}
\newcommand{\rmnum}[1]{\romannumeral #1}
\newcommand{\Rmnum}[1]{\uppercase\expandafter{\romannumeral #1}}

\title{Quantization of the electromagnetic field outside high-dimensional static black holes and its application outside the Gauss-Bonnet black hole}

\author{Ming Zhang$^{1,2}$, Zhan-Ying Yang$^{2}$, Rui-Hong Yue$^{1}$\footnote{yueruihong@nbu.edu.cn}}

\address{$^{1}$Faculty of Science, Ningbo University, Ningbo 315211, China\\
$^{2}$Department of Physics, Northwest University, Xi'an, 710069, China}

\begin{abstract}
  In present paper, we investigated the quantization of an electromagnetic field in the background  of static spherically symmetric $d$-dimensional spacetime in the Boulware vacuum. We have also calculated the response rate of a static charge outside both $d$-dimensional Schwarzschild black hole and the Gauss-Bonnet black hole in the low-frequency regime, which can be expressed as the summation of hypergeometric functions.
\end{abstract}


\maketitle

\section{INTRODUCTION}

At present it is believed that the study of quantum field theory in curved spacetime can provide some insights into quantum gravity effects, while the full theory is not available.
An important prediction in this field is the thermal evaporation of black hole \cite{Hawking:1974sw}. This nontrivial effect was soon realized to be closely associated with the existence of an event horizon in Schwarzschild spacetime.
One of the difficulties in studying fields in Schwarzschild \cite{Jensen:1985in} and other black hole spacetime, even when the fields are non-interacting, stems from the fact that the solutions to the field equations are functions whose properties are not well known. In the low-frequency regime, however, the situation is much simpler and the mode functions of the massless scalar field are well known \cite{Candelas:1980zt}. Recently, Crispino et al. \cite{Crispino:2000jx} suggested a scheme to quantize  the free quantum electrodynamics in static  spherically symmetric $d$-dimensional spacetime and gave out the response rate of a static charge outside the four-dimensional Schwarzschild black hole.

Following the advent of string theory, extra dimensions were promoted from an interesting curiosity to a theoretical necessity since superstring theory requires a ten-dimensional spacetime to be consistent from a quantum point of view (\cite{Horava:1996ma}-\cite{Randall:1999vf}). Among the higher curvature gravities, the most extensively studied theory is the so-called Gauss-Bonnet gravity (\cite{Lanczos:1938sf} -\cite{Wheeler:1985qd}), which naturally emerges when we want to generalize Einstein's theory in higher dimensions by keeping all characteristics of usual general relativity excepting the linear dependence of the Riemann tensor. Therefore, it is necessary to study the quantization of the free electromagnetic field with the high-dimensional spacetime and its specific application outside a Gauss-Bonnet black hole.

In this paper we examine free quantum electrodynamics in static spherically symmetric spacetime of arbitrary dimensions in a modified Feynman gauge \cite{Crispino:2000jx}. We give all of the physical modes functions which consummate the results in Ref.\cite{Crispino:2000jx}. Then we calculate the response rate (The response rate is a quantum concept with no natural analog in classical physics though it is possible to define a corresponding classical quantity mathematically and represents  the number of times the source responds to the field per unit time.) of a static charge outside the $d$-dimensional Schwarzschild black hole and the $d$-dimensional GB black hole in the Unruh vacuum \cite{Unruh:1976db}. Limited  to four-dimensional Schwarzschild black hole, the response rate is consistent with the result in Ref.\cite{Crispino:2000jx}.

The paper is organized as follows. In Sec.II we review basic concepts of the electromagnetic field in the arbitrary dimensional spacetime of a spherically symmetric black hole in a modified Feynann gauge. Then we show how the Gupta-Bleuler condition is implemented to obtain the physical states. Sec.III and Sec.IV devote to calculate the response rate of a static charge outside a Schwarzschild black hole and a GB black hole of arbitrary-dimensions respectively. In Sec.V, we summarize the main results. And a brief proof about our solutions of physical modes are given in Appendix.

\section{GUPTA-BLEULER QUANTIZATION IN A MODIFIED FEYNMAN GAUGE}

In this section, we follow the notation of Ref.\cite{Crispino:2000jx} to study the solutions of field equations for electromagnetic field in an asymptotic flat and static spherically symmetric $(p+2)$-dimensional spacetime. The quantization of electromagnetic field will be carried out in the frame of Gupta-Bleuler formalism in a modified Feynmann gauge.

The line element under considered takes the form
\begin{equation}
  d\tau^2=f(r)dt^2-h(r)dr^2-r^2ds_p^2
\end{equation}
 with the line element of a unit $p$-sphere $ds_p^2$. We assume that both $f(r)$ and $h(r)^{-1}$ have a zero at $r=r_h$ and  positive for $r>r_h$.

The  Lagrangian density for electromagnetic field in a modified Feynman gauge is
\begin{equation}
  \mathcal{L}_F=\sqrt{-g}[-\frac{1}{4}F_{\mu\nu}F^{\mu\nu}-\frac{1}{2}G^2],\label{lagEM}
\end{equation}
and $G$ stands for the modified Feynman gauge
\begin{equation}
  \label{equ:g}
  G=\nabla^{\mu}A_{\mu}+K^{\mu}A_{\mu}.
\end{equation}
Here the vector $K^{\mu}$ is independent on electromagnetic field $A_\mu$, and takes the form
\begin{equation}
  \label{equ:k}
  K^{\mu}=(0,f'/(fh),0,0).
\end{equation}
Under this choice, the gauge condition changes into
\begin{equation}
  G=\frac{1}{f}\partial_t A_t-\sqrt{\frac{f}{h}}\frac{1}{r^p}\partial_r[\frac{r^p}{\sqrt{fh}}A_r]-\frac{1}{r^2}\nabla^i A_i
\end{equation}
From the  Lagrangian density for electromagnetic field, the equations of motion  are
\begin{subequations}
  \label{equ:mm}
  \begin{numcases}{}
  -\frac{1}{f}\partial_t^2 A_t+\sqrt{\frac{f}{h}}\frac{1}{r^p}\partial_r\Big[\frac{r_p}{\sqrt{fh}}\partial_r A_t\Big]+\frac{1}{r^2}\nabla^2 A_t=0 \\
  -\frac{1}{f}\partial_t^2 A_r+\frac{1}{f}\partial_r\Big[\sqrt{\frac{f}{h}}\frac{f}{r^p}\partial_r\big(\frac{r_p}{\sqrt{fh}} A_r\big)\Big]+\frac{1}{r^2}\nabla^2 A_r+\frac{1}{f}\partial_r\big(\frac{f}{r^2}\big)\nabla^i A_i=0 \\
  -\frac{1}{f}\partial_t^2 A_i+\frac{r^{2-p}}{\sqrt{fh}}\partial_r\Big(\sqrt{\frac{f}{h}}r^{p-2}\partial_r A_i\Big)-\frac{r^2}{fh}\partial_r\Big(\frac{f}{r^2}\Big)\partial_i A_r \nonumber\\
 \quad +\frac{1}{r^2}\Big[\nabla^j\big(\nabla_j A_i-\nabla_i A_j\big)+\partial_i\big(\nabla^j A_j\big)\Big]=0,\qquad (j=1\cdots p).
  \end{numcases}
\end{subequations}
Here $\nabla_i$ is the  covariant derivative on $S^p$.

We  denote the  complete set of solutions of \eq{equ:mm}  by $A_{\mu}^{(\lambda n;\omega lm)}$, and  call a non-physical modes for  $\lambda=0$, physical modes for $\lambda=1,2,3,...,p$ and a pure-gauge mode for $\lambda=p+1$. The label $n$ represents modes incoming from the past null infinity $(n=\leftarrow)$ and coming out from the past horizon $(n=\rightarrow)$ .

\subsection{NON-PHYSICAL AND PURE-GAUGE MODES}

The non-physical modes is a static electric field solution of \eq{equ:mm} with a gauge condition $G=0$. In such mode all components  of $A_\mu$  are zero excepting for $A_t$ \cite{Crispino:2000jx}
\begin{subequations}
  \begin{numcases}{}
    A_t^{(0n;\omega lm)}=R_{\omega l}^{(0n)}(r)Y_{lm}e^{-i\omega t}\\
    A_{\mu}^{(0n;\omega lm)}=0, ~~(\mu\neq t).
  \end{numcases}
\end{subequations}
Here $Y_{lm}$ is a scalar spherical harmonic function on the unit $p$-sphere \cite{Chodos:1983zi} satisfying
\begin{eqnarray}
  \nabla^2 Y_{lm}=-l(l+p-1)Y_{lm} \label{equ:y}
\end{eqnarray}
where $l=0,1,2,...$ and $m$ denote a set of $p-1$ integers $m_1,...m_{p-1}$ satisfying $l\geq m_{p-1}\geq ...\geq m_2 \geq\mid m_1\mid$. They are normalized as
\begin{equation}
  \int\ud \Omega_p\overline{Y_{lm}} Y_{l'm'}=\delta_{ll'}\delta_{mm'}.
\end{equation}
The radius function $R_{\omega l}^{(0n)}(r)$ is governed by
\begin{equation}
  \Big[\frac{\omega^2}{f}+\sqrt{\frac{f}{h}}\frac{1}{r^p}\frac{\ud}{\ud r}\big(\frac{r^p}{\sqrt{fh}}\frac{\ud}{\ud r}\big)-\frac{l(l+p-1)}{r^2}\Big]R_{\omega l}^{(0n)}(r)=0.
\end{equation}


The pure-gauge mode is given  by \cite{Crispino:2000jx}
\begin{subequations}
  \begin{numcases}{}
    A_{\mu}^{((p+1) n;\omega lm)}=\nabla_{\mu}\Lambda^{(n\omega lm)},  \\
    \Lambda^{(n\omega lm)}=\frac{i}{\omega}R_{\omega l}^{(0n)}(r)Y_{lm}e^{-i\omega t}
  \end{numcases}
\end{subequations}
which satisfy the field equations \eq{equ:mm} and $G=0$. (In Ref.\cite{Crispino:2000jx}, $p+1=3$.)

\subsection{PHYSICAL MODES}

For other independent solutions $\lambda=1,2,...,p$, which represent physical degrees of freedom, the time-component can be taken as zero. They are the  linear independent solution of \eq{equ:mm} with a gauge condition $G=0$, and are classified into two types.

\subsubsection{PHYSICAL MODES \Rmnum{1}}

The ``physical modes \Rmnum{1}'' solution can be written as \cite{Crispino:2000jx}
\begin{subequations}
  \begin{numcases}{}
    A_t^{(1n;\omega lm)}=0 \\
    A_r^{(1n;\omega lm)}=R_{\omega l}^{(1n)}(r)Y_{lm}e^{-i\omega t}, (l\geq1)\\
    A_i^{(1n;\omega lm)}=\frac{r^{2-p}}{l(l+p-1)}\sqrt{\frac{f}{h}}\frac{\ud}{\ud r}\Big[\frac{r^p}{\sqrt{fh}}R_{\omega l}^{(1n)}(r)\Big]\partial_i Y_{lm}e^{-i\omega t}
  \end{numcases}
\end{subequations}
where $i=1,2,3,...,p$, $Y_{lm}$ has been declared in \eq{equ:y} and $R_{\omega l}^{(1n)}(r)$ is governed by
\begin{equation}
  \label{equ:re}
  \Big[\frac{\omega^2}{f}-\frac{l(l+p-1)}{r^2}\Big]R_{\omega l}^{(1n)}(r)+\frac{1}{r^2}\frac{\ud}{\ud r}\Big[\sqrt{\frac{f}{h}} r^{2-p}\frac{\ud}{\ud r}\big(\frac{r^p}{\sqrt{fh}}R_{\omega l}^{(1n)}(r)\big)\Big]=0
\end{equation}

\subsubsection{PHYSICAL MODES  \Rmnum{2}}

The solutions of the ``physical modes \Rmnum{2}'' $A_{\mu}^{(sn;\omega lm)}$ can be written $(s=2,3,...,p)$
\begin{subequations}
  \label{equ:as}
  \begin{numcases}{}
    A_t^{(sn;\omega lm)}=A_r^{(sn;\omega lm)}=A_i^{(sn;\omega lm)}=0 , (1\le i\le s-2)\\
    A_i^{(sn;\omega lm)}=R_{\omega l}^{(sn)}\Phi_i^{(s;lm)}(p) e^{-i\omega t}, (s-1\le i \le p)
  \end{numcases}
\end{subequations}
and
\begin{eqnarray}
  \Big[\frac{\omega^2}{f}+\frac{1}{\sqrt{fh} r^{p-2}}\frac{\ud}{\ud r}\big(\sqrt{\frac{f}{h}}r^{p-2}\frac{\ud}{\ud r}\big)\Big]R_{\omega l}^{(sn)}(r)
   -\frac{(L_p+1)(L_p+p-2)}{r^2}R_{\omega l}^{(sn)}(r)=0
\end{eqnarray}

The functions  $\Phi_i^{(s;lm)}(p)$ should satisfy the following equation
\begin{subequations}
  \label{equ:phyr}
  \begin{numcases}{}
  \nabla^k(\nabla_k \Phi_i^{(s;lm)}-\nabla_i \Phi_k^{(s;lm)})=0 ~,~(1\leq i\leq s-2) \\
  \nabla^k(\nabla_k \Phi_i^{(s;lm)}-\nabla_i \Phi_k^{(s;lm)})=-(L_p+1)(L_p+p-2)\Phi_i^{(s;lm)} ~,~ (s-2<i\leq p) \label{equ:phy2},\\
  L_p=L-\frac{p-2}{2},
  \end{numcases}
\end{subequations}
where $L$ is an arbitrary number which satisfies the relation $(L\ge l+\frac{p-2}{2})$. The group number s ``$L_p,~L_{p-1},...,L_{p-s+3}$'' we constructed make the function $\Phi^{(s;lm)}$ meet the completeness conditions.
For 'Physical Modes \Rmnum{2}', the super index $(lm)$ in $A^{(sn,\omega lm)}$, $\Phi^{(s;lm)}$ and $R_{\omega l}^{(sn)}$  stands for a set of quantities $$(L_p,\cdots,L_{p-s+3},l,m_{p-s+1},\cdots,m_1),$$ and the last $(p-s+2)$ parameters $(l,{m_i})$ are integers.

The solutions of \eq{equ:phy2} are
\begin{subequations}
  \label{equ:phy}
  \begin{numcases}{}
    \Phi_i^{(s;lm)}(p)=0 ~,~ (1\leq i\leq s-2) \\
    \Phi_i^{(s;lm)}(p)=F^s(\theta)\widetilde{Y_i}^{(lm)}(N=p-s+2) ~,~ (s-2<i\leq p) \\
    F^s(\theta)=\prod_{j=p+3-s}^p\sin^{\frac{4-j}{2}}({\theta_{p-j+1}})\mathrm P_{L_j+\frac{j-2}{2}}^{L_{j-1}+\frac{j-2}{2}}(\cos{\theta_{p-j+1}})
  \end{numcases}
\end{subequations}
Here $\mathrm P_a^b(\theta)$ is the associated Legendre function and $\widetilde{Y_i}^{(lm)}(N)$ are the divergence-free vector spherical harmonics on the unit $N$-sphere satisfying
\begin{eqnarray}
  \label{equ:ym}
  &&\widetilde{\nabla}^k(\widetilde{\nabla}_k \widetilde{Y_i}^{(lm)}(N)-\widetilde{\nabla}_i \widetilde{Y_k}^{(lm)}(N))=-(L_N+1)(L_N+N-2)\widetilde{Y_i}^{(lm)}(N)
\end{eqnarray}
with the notation  $\widetilde{\nabla}_i$ as the associated covariant derivative on $S^N$. The  solutions of \eq{equ:ym} are
\begin{subequations}
  \begin{numcases}{}
  \widetilde{Y}_{p-N+1}^{(lm)}(N)=H(\theta) Y_{lm}(\theta_{p-N+2},...,\theta_p) ; \label{equ:yy}\\
  \widetilde{Y_i}^{(lm)}(N)=\frac{\sin^2(\theta_{p-N+1})}{l(l+N-2)}\partial_i Y_{lm}(\theta_{p-N+2},...,\theta_p) \nonumber\\
  \times(\partial_{\theta_{p-N+1}}+(N-1)\cot{(\theta_{p-N+1})}) H(\theta) ;~(p-N+1<i\leq p )\\
  H(\theta)=\sin{(\theta_{p-N+1})}^{-\frac{N}{2}}P_{L_N+\frac{N-2}{2}}^{l+\frac{N-2}{2}}(\cos{\theta_{p-N+1}})
  \end{numcases}
\end{subequations}
Here $Y_{lm}(\theta_{p-N+2},...,\theta_p)$ is a scalar spherical harmonic on the unit (N-1)-sphere. The proof  will be  given in Appendix A.

It is clear  that the number of "physical modes  \Rmnum{2}" is $(p-1)$, which (including physical modes \Rmnum{1})is equivalent with the degree of the $p$-sphere. In the Ref.\cite{Crispino:2000jx}, the physical modes \Rmnum{2} has been studied, but there is only one solution, which is equal to the \eq{equ:as} with  $s=2$. Thus the solutions \eq{equ:as} with  $3\le s\le p$ are new and necessary for the completeness of solution.

\subsection{QUANTIZATION}

Using Gupta-Bleuler quantization, we impose the equal-time commutation relations on the field $\hat{A}_{\mu}$ and momentum $\hat{\Pi}^{t\mu}$ operators
\begin{equation}
  [\hat{A}_{\mu}(t,\textbf{x}),\hat{A}_{\nu}(t,\textbf{x}')]=[\hat{\Pi}^{t\mu}(t,\textbf{x}),\hat{\Pi}^{t\nu}(t,\textbf{x}')]=0
\end{equation}
\begin{equation}
  [\hat{A}_{\mu}(t,\textbf{x}),\hat{\Pi}^{t\nu}(t,\textbf{x}')]=\frac{i\delta_{\mu}^{\nu}}{\sqrt{-g}}\delta^{p+1}(\textbf{x}-\textbf{x}')
\end{equation}
where $\textbf{x}$ and $\textbf{x}'$ represent all spatial coordinates.
The field $\hat{A}_{\mu}$ can be expressed in terms of  $A_{\mu}^{(\xi)}$
\begin{equation}
  \hat{A}_{\mu}(t,\textbf{x})=\sum_{\rho}\int_{-\infty}^{+\infty}\frac{\ud \omega}{\sqrt{4\pi|\omega|}}A_{\mu}^{(\omega\rho)}(t,\textbf{x})a_{\omega\rho},
\end{equation}
where $A_{\mu}^{(-\omega\rho)}\equiv \overline{A_{\mu}^{(\omega\rho)}}$ , $a_{-\omega\rho}\equiv a_{\omega\rho}^{\dag}$ and $\rho$ labels all quantum numbers.

\section{RESPONSE RATE OF A STATIC CHARGE OUTSIDE A D-DIMENSIONAL SCHWARZSCHILD BLACK HOLE}

 In this section, we will calculate  the response rate of a static charge outside a $d$-dimensional $(d=p+2)$ Schwarzschild black hole by following the procedure of Ref.\cite{Crispino:2000jx}. In this case, the black hole is characterized by  $f(r)=h(r)^{-1}=1-(r_h/r)^{(p-1)}$.

In order to avoid the indefinite results (\cite{Higuchi:1992we},\cite{Higuchi:1992td}), we use the formula suggested by  Crispino et al. \cite{Crispino:1998hp} and  assume  the static charge located at $(r_0,\theta_{0})$ with a current density  $j^{\mu}$
\begin{subequations}
  \label{equ:js}
  \begin{numcases}{}
    j^{\mu}=(j^{t},j^{r},0,...,0) \label{equ:j} \\
    j^{t}=\frac{\sqrt{2}q\cos{Et}}{\sqrt{-g}}\delta(r-r_0)\delta(\theta_1-\theta_{10})\cdot...\cdot\delta(\theta_p-\theta_{p0}) \\
    j^{r}=\frac{\sqrt{2}qE\sin{Et}}{\sqrt{-g}}\Theta(r-r_0)\delta(\theta_1-\theta_{10})\cdot...\cdot\delta(\theta_p-\theta_{p0})
  \end{numcases}
\end{subequations}
The step function $\Theta(x)$ is defined by $\Theta(x)=1,(x>0)$ and vanishing for $x\le 0$. We take the limit $E\rightarrow 0$ in the end, assuming that the rate is continuous at $E=0$.

Such current interacts with vector potential $A_\mu$ through the Lagrangian $\sqrt{-g}j^\mu A_\mu$. Since  $a_{\omega lm}^{((p+1)n)\dag}|\text{phys}>$ is non-physical states,  which excludes the interaction with the  pure-gauge particles created by $a_{\omega lm}^{((p+1)n)\dag}$. We will neglect it. However, the current does interact with the states created by $a_{\omega lm}^{(0n)\dag}$ but the contribution to physical probabilities maybe taken as zero once the non-physical modes are appropriately chosen. Furthermore, there is no interacting  term  between the current and physical modes \Rmnum{2} due to $A_t=A_r=0$. Therefore,  we only need to consider the physical modes \Rmnum{1}.
To  make the counting process more concise, we will limit in  the  spherical coulomb gauge. The mode function can be written as \cite{Crispino:2000jx}
\begin{eqnarray}
  &&A_t^{1'n;\omega lm}=\frac{i\omega r^{2-p}}{l(l+p-1)}\sqrt{\frac{f}{h}}\frac{\ud}{\ud r}(\frac{r^p}{\sqrt{fh}}R_{\omega l}^{1n})Y_{lm}e^{-i\omega t} \\
  &&A_r^{1'n;\omega lm}=\frac{\omega^2 r^2}{l(l+p-1)}\frac{1}{f}R_{\omega l}^{1n}Y_{lm}e^{-i\omega t}.
\end{eqnarray}

In the limit $E\rightarrow 0$, the proper response rate of the charge can be written \cite{Crispino:2000jx}
\begin{eqnarray}
  \label{equ:r0}
  \frac{R_{0lm}}{\sqrt{f(r_0)}}=4\pi\lim_{E\rightarrow 0}\frac{|\mathcal{T}_{Elm}^{\rightarrow}|^2}{\sqrt{f(r_0)}\beta E}
\end{eqnarray}
where $\mathcal{T}_{Elm}^{\rightarrow}$ is the index $n=\rightarrow$ in transition amplitude $\mathcal{T}_{\omega lm}^n$ which has the form
\begin{eqnarray}
  \label{equ:a}
  \mathcal{T}_{\omega lm}^n\equiv \frac{1}{2\pi \delta(\omega-E)}\int\ud^{p+2} x~\sqrt{-g}j^{\mu}\langle 1n;\omega lm|\hat{A}_{\mu}|0\rangle
\end{eqnarray}

Now, we return to  calculate $\mathcal{T}_{Elm}^{\rightarrow}$. The \eq{equ:re} can be written
\begin{equation}
  \label{equ:radius}
  \frac{1}{r^2}\frac{\ud}{\ud r}[fr^{2-p}\frac{\ud}{\ud r}(r^{p}R_{\omega l}^{(1\rightarrow)}(r))]-\frac{l(l+p-1)}{r^2}R_{\omega l}^{(1\rightarrow)}(r)+\frac{\omega^2}{f}R_{\omega l}^{(1\rightarrow)}(r)=0.
\end{equation}
After introducing the Wheeler tortoise coordinate and a function $\varphi$
\begin{equation}
  R_{\omega l}^{(1\rightarrow)}(r)\equiv \frac{\sqrt{l(l+p-1)}}{\omega}r^{-\frac{p}{2}-1}\varphi_{\omega l}^{(1\rightarrow)}(r),
\end{equation}
the  \eq{equ:radius} changes into
\begin{equation}
  (\omega^2+\frac{\ud^2}{\ud r^{*2}}-V_1(r^*))\varphi_{\omega l}^{(1\rightarrow)}(r)=0
\end{equation}
with
\begin{eqnarray}
    &&V_1[r^*(r)]=f\frac{l(l+p-1)}{r^2}+f^2\frac{p(p-2)}{4r^2}-f'f\frac{(p-2)}{2r}
\end{eqnarray}

For the small $\omega$ and the condition $(r-r_h\ll \omega^2r_h^3, |\omega r^*|\ll 1)$, the wave coming from the past horizon $H^-$ is almost completely reflected back by the potential toward the horizon
\begin{equation}
  \label{equ:v}
  \varphi_{\omega l}^{(1\rightarrow)}\approx -2\omega r^*+\text{const}
\end{equation}

Generally, it is hard to find the analytic expression for the  Wheeler tortoise coordinate $r^*$. Fortunately, what we need is just the behavior of $r^*$ near horizon. Substituting the expression of $f(r)$, the leading term of the Wheeler tortoise coordinate can be written
\begin{equation}
  r^* \approx \frac{r_h}{p-1}\ln{(z-1)},(z\sim 1)
\end{equation}
by using the transition
\begin{equation}
  z=2(\frac{r}{r_h})^{p-1}-1
\end{equation}

Thus,  the boundary condition of $R_{\omega l}^{(1\rightarrow)}$ reads
\begin{equation}
  \label{equ:bc}
  R_{\omega l}^{(1\rightarrow)} \approx -\frac{2\sqrt{l(l+p-1)}}{p-1}2^{\frac{1}{p-1}}M^{-\frac{p}{2(p-1)}}\ln{(z-1)}, ~~(r-r_h\ll \omega^2r_h^3, |\omega r^*|\ll 1),
\end{equation}
which is independent on $\omega$. In terms of variable $z$, the \eq{equ:radius} can be written as
\begin{eqnarray}
   \label{equ:rz}
   &&\displaystyle (z^2-1)\frac{\ud^2 R_{\omega l}^{(1\rightarrow)}}{\ud z^2}+2\left[\frac{p}{p-1}(z-1)+1\right]\frac{\ud R_{\omega l}^{(1\rightarrow)}}{\ud z}
   -\frac{1}{(p-1)^2}\Big[l(l+p-1)\nonumber \\ &&\qquad \left.-\frac{2p(p-2)}{z+1}-p-\omega^2\frac{z+1}{z-1}(M(z+1))^{\frac{2}{p-1}}\right]R_{\omega l}^{(1\rightarrow)}=0 .\label{equ:hyper1}
\end{eqnarray}
This equation can not be analytically  solved, but the boundary condition implies that the main contribution is from the $\omega$-independent term.   Combining the  asymptotic behavior $R_{\omega l}^{(1\rightarrow)} \rightarrow 0$ as $z \rightarrow +\infty$,   we find the solution of \eq{equ:rz} in small $\omega$ limit
\begin{equation}
  \label{equ:r11}
  R_{\omega l}^{(1\rightarrow)}=\frac{2\sqrt{l(l+p-1)}}{p-1}2^{\frac{1}{p-1}}M^{-\frac{p}{2(p-1)}} \mathrm F([\frac{l}{p-1},\frac{l}{p-1}+2],[\frac{2l}{p-1}+2],
  \frac{2}{z+1})(z+1)^{\frac{-l-p}{p-1}}
\end{equation}
where $\mathrm F(\alpha,\beta,\gamma,x)$ is the hypergeometric function and the coefficient has been appropriately chosen to agreement with the boundary condition \eq{equ:bc}.

From gauge invariance we can see that it is convenient to calculate the amplitude \eq{equ:a} in the spherical Coulomb gauge. Then the contribution from the $r$-component will be suppressed in the low energy limit due to extra factors of $\omega$. So we only need to consider the $t$-component in this limit.
By using the definition of \eq{equ:r0}, we can get
\begin{equation}
  \frac{R_{0lm}}{\sqrt{f(r_0)}}=\frac{2^{\frac{2-p}{p-1}}(p-1) q^2}{M\pi l(l+p-1)\sqrt{f(r_0)}}(\frac{z_0-1}{z_0+1})^2[\frac{\ud}{\ud z_0}(\mathrm F_l(z_0)(z_0+1)^{\frac{-l}{p-1}})]^2 |Y_{lm}|^2
\end{equation}

Substituting the all parameters for $d$-dimensional black hole
\begin{eqnarray}
  &&z_0=\frac{r_0^{p-1}}{M}-1 \\
  &&f(r_0)=1-\frac{2M}{r_0^{p-1}} \\
  &&\mathrm F_l(z_0)=\mathrm F([\frac{l}{p-1},\frac{l}{p-1}+2],[\frac{2l}{p-1}+2],
  \frac{2}{z_0+1})
\end{eqnarray}
the total transition probability per proper time of the charge is given by
\begin{eqnarray}\label{pro-schwarzschild}
  \label{equ:rr1}
  &&R^{\text{tot}}=\sum_l^{} \sum_{m=-1}^l\frac{R_{0lm}}{\sqrt{f(r_0)}}  \nonumber\\
  &&=\sum_l^{} \frac{2^{\frac{2-p}{p-1}} (p-1) q^2}{M\pi l(l+p-1)\sqrt{f(r_0)}}(\frac{z_0-1}{z_0+1})^2[\frac{\ud}{\ud z_0}(\mathrm F_l(z_0)(z_0+1)^{\frac{-l}{p-1}})]^2 \frac{G(l)}{\Omega_{p}}
\end{eqnarray}
in which the following formula has been used \cite{Camporesi:1994ga}
\begin{eqnarray}
  \sum_{m=-1}^l |Y_{lm}|^2=\frac{G(l)}{\Omega_{p}}
\end{eqnarray}
where $\Omega_p$ is the volume of $S^{p}$ and $G(l)$ is the degeneracy of the eigenvalue $-l(l+p-2)$ of the Laplacian $\tilde{\Delta}$, which is given by
\begin{equation}
  G(l)=\frac{(2l+p-1)(l+p-2)!}{l!(p-1)!}.
\end{equation}

When the dimension reduced to four ($p=2$), \eq{equ:r11} can be written
\begin{equation}
  R_{\omega l}^{(1\rightarrow)}=\frac{4}{M}\sqrt{l(l+1)}\mathrm F([l,l+2],[2l+2],\frac{2}{z+1})(z+1)^{-l-2},
\end{equation}
and \eq{pro-schwarzschild} will reduce to a Legendre function of the second kind, which recovers the result in \cite{Crispino:2000jx} .

\section{RESPONSE RATE OF A STATIC CHARGE OUTSIDE A D-DIMENSIONAL GAUSS-BONNET GRAVITY BLACK HOLE}

The action of pure  GB can be written as
\begin{equation}
  S=\int\ud^{p+2}x~\sqrt{-g}\alpha \mathcal {L}_2
\end{equation}
where the coupling constant $\alpha$ can be regarded as the inverse of string tension and be assumed $\alpha>0$ in this paper.
The Gauss-Bonnet term is given by
\begin{equation}
  \mathcal {L}_2=R^2-4R_{ab}R^{ab}+R_{abcd}R^{abcd}.
\end{equation}

The line element for the exterior region of the $d$-dimensional Gauss-Bonnet gravity is given by
\begin{equation}
  \ud s^2=f(r)\ud t^2-f(r)^{-1}\ud r^2-r^2\ud s_p^2
\end{equation}
The $f(r)$ in the $d$-dimensional GB gravity is
\begin{equation}
  f(r)=1-\sqrt{\frac{2M}{\alpha r^{d-5}}}\equiv 1-\frac{B}{r^{\frac{d-5}{2}}},
\end{equation}
The horizon radius $r_h=(\frac{2M}{\alpha})^{\frac{1}{d-5}}=B^{\frac{2}{d-5}}$.

By using the same assumption and conclusion \eq{equ:js} to \eq{equ:v}, we can continue to compute the response rate of the static electric charge interacting with photons of Hawking radiation in the Unruh vacuum.

We consider the behavior of $r^*$ near horizon, the leading term of the Wheeler tortoise coordinate can be written
\begin{equation}
  r^* \approx \frac{2}{p-3}(\frac{2M}{\alpha})^\frac{1}{p-3}\ln{(z-1)}
\end{equation}
by using the transition
\begin{equation}
  z=\frac{2}{B}r^{\frac{p-3}{2}}-1.
\end{equation}

The boundary condition of $R_{\omega l}^{(1\rightarrow)}$ reads
\begin{equation}
  \label{equ:bc2}
  R_{\omega l}^{(1\rightarrow)} \approx -\frac{\sqrt{l(l+p-1)}}{p-3}2^{\frac{3p-4}{p-3}} B^{-\frac{p}{p-3}}\ln{(z-1)}, ~~(r-r_H\ll \omega^2r_H^3, |\omega r^*|\ll 1)
\end{equation}
In terms of variable $z$, \eq{equ:radius} can be written
\begin{eqnarray}
   &&\frac{1}{4}(z^2-1)\frac{\ud^2 q(z)}{\ud z^2}+\frac{1}{4}[\frac{3p-1}{p-3}(z-1)+2]\frac{\ud q(z)}{\ud z}  \nonumber\\
   &&+\frac{1}{(p-3)^2}[-l(l+p-1)+\frac{p(p-5)}{z+1}+p+\omega^2\frac{z+1}{z-1}(\frac{B}{2}(z+1))^{\frac{4}{p-3}}]q(z)=0 \label{equ:hyper2}
\end{eqnarray}
and this equation can be solved explicitly for small $\omega$ limit. Combining the asymptotic behavior $R_{\omega l}^{(1\rightarrow)} \rightarrow 0$ as $z \rightarrow +\infty$, the solution of \eq{equ:hyper2} is
\begin{equation}
  R_{\omega l}^{(1\rightarrow)}=\frac{\sqrt{l(l+p-1)}}{p-3}2^{\frac{3p-4}{p-3}}B^{-\frac{p}{p-3}} \mathrm F([\frac{2l}{p-3},\frac{-2l-2p+2}{p-3}],[\frac{-2p+2}{p-3}],\frac{z+1}{2})
  (z+1)^{-\frac{2p}{p-3}}
\end{equation}
where the coefficient has been appropriately chosen to agreement with the boundary condition \eq{equ:bc2}.

Thus, the expression of the proper response rate of the charge is
\begin{equation}
  \frac{R_{0lm}}{\sqrt{f(r_0)}}=\frac{(p-3)q^2(z_0+1)^{-\frac{2p+2}{p-3}}B^{\frac{2-2p}{p-3}} 2^{\frac{6}{p-3}}}{\pi l(l+p-1)\sqrt{f(r_0)}}(\frac{z_0-1}{z_0+1})^2[\frac{\ud}{\ud z_0}\mathrm F_l(z_0)]^2|Y_{lm}|^2
\end{equation}
Substituting the all parameters for $d$-dimensional GB black hole
\begin{eqnarray}
  &&z_0=\frac{2}{B}r_0^{\frac{p-3}{2}}-1 \\
  &&f(r_0)=1-\frac{B}{r^{\frac{d-5}{2}}_0} \\
  &&\mathrm F_l(z_0)=\mathrm F([\frac{2l}{p-3},\frac{-2l-2p+2}{p-3}],[\frac{-2p+2}{p-3}],\frac{z+1}{2})
\end{eqnarray}
the total transition probability per proper time of the charge is given by
\begin{eqnarray}
  \label{equ:rr2}
  &&R^{\text{tot}}=\sum_l^{+\infty} \sum_{m=-1}^l\frac{R_{0lm}}{\sqrt{f(r_0)}}  \nonumber\\
  &&=\sum_l^{+\infty} \frac{(p-3)q^2(z_0+1)^{-\frac{2p+2}{p-3}}B^{\frac{2-2p}{p-3}} 2^{\frac{6}{p-3}}}{\pi l(l+p-1)\sqrt{f(r_0)}}(\frac{z_0-1}{z_0+1})^2[\frac{\ud}{\ud z_0}\mathrm F_l(z_0)]^2 \frac{G(l)}{\Omega_{p}}
\end{eqnarray}

We can give the main contribution of the result of \eq{equ:rr1} and \eq{equ:rr2},
\begin{eqnarray}
  &&R^{\text{tot}}\sim \big(\frac{r_h}{r_0}\big)^{p-3}, ~(\text{Schwarzschild}) \label{equ:rs} \\
  &&R^{\text{tot}}\sim \big(\frac{r_h}{r_0}\big)^{\frac{3p-1}{2}}.~(\text{GB}) \label{equ:rg}
\end{eqnarray}
For the case of $d\ge 5$, assuming that there is a charge outside an unknown black hole, the radius of the horizon ($r_h$), the mass of the black hole ($M$) and the location of the charge ($r_0$) are known. Meanwhile, we can measure the response rate of the charge outside this unknown black hole. Compare the measured value with \eq{equ:rs} and \eq{equ:rg}, then we can tell the black hole is the Schwarzschild type or the GB type.

\section{CONCLUSIONS}

In this paper we quantized the free electrodynamics in static spherically symmetric spacetime of arbitrary dimensions in a modified Feynman gauge. Then we examined the Gupta-Bleuler quantization in this modified Feynman gauge. The results obtained were applied to compute the total response rate of a static charge outside the $d$-dimensional Schwarzschild black hole and the $d$-dimensional  GB black hole in the Unruh vacuum.

For the Einstein-Gauss-Bonnet gravity one can follow the same procedure. The  Wheeler tortoise coordinate $r^*$ has a asymptotic behavior $r^*\approx f'(r_h)^{-1}\ln(r-r_h)$ as near horizon. The boundary condition \eq{equ:v} changes into $ \varphi_{\omega l}^{(1\rightarrow)}\approx -2\omega/f'(r_h)\ln{(r-r_h)} $. One problem is that we do not find a new variable to simplify the \eq{equ:radius} and its analytic solution.  An applicable way is to find a series solution of \eq{equ:radius}, the results can not be expressed in terms of familiar special functions and we neglect it here.

Such an outcome is not only a simple promotion work for what Crispino et al. \cite{Crispino:2000jx} have done. Having the specific form of the free quantum electrodynamics in static spherically symmetric spacetime of arbitrary dimensions, It may provide us a chance for further investigating quantum field theory in high-dimensional curved spacetime. For instance, some authors (\cite{Crispino:1998hp}, \cite{Candelas:1983gz}, \cite{Kleinert:1996tm}) have researched whether or not a quantum version of the equivalence principle could be formulated and show some equivalence for low-frequency quantum phenomena in flat and curved spacetime. The same problem could be reconsidered  in high-dimensional spacetime and discuss the dimension dependence of the results.

\appendix
\section{THE VERIFICATION OF PHYSICAL MODES \Rmnum{2}}

For concise sake, we use $\Phi_{\mu}^{(p)}$ to denote $\Phi_{\mu}^{(s;lm)}(p)$, and let
\begin{subequations}
  \label{equ:1}
  \begin{numcases}{}
    \Phi_1^{(p)}=0 \\
    \Phi_i^{(p)}=\sin^{\frac{4-p}{2}}{(\theta_1)}\mathrm P_{L_p+\frac{p-2}{2}}^{L_{p-1}+\frac{p-2}{2}}(\cos{\theta_1})\Phi_i^{(p-1)},~~(2\le i \le p).
  \end{numcases}
\end{subequations}
In order to verify that $\Phi_{\mu}^{(p)}$ satisfy the relations \eq{equ:phyr}, we need to give the following relation
\begin{subequations}\label{equ:2}
  \begin{numcases}{}
    \nabla^{(p)\mu}\nabla_{\mu}^{(p)} \Phi_{1}^{(p)}=\frac{1}{\sin^2{\theta_1}}\nabla^{(p-1)k}\nabla_k^{(p-1)} \Phi_{1}^{(p)}+\partial_1^2\Phi_{1}^{(p)}+(p-1)\cot{\theta_1}\partial_1\Phi_{1}^{(p)} \nonumber\\
   \qquad\qquad\qquad-(p-1)\cot^2{\theta_1}\Phi_{1}^{(p)}-2\cot{\theta_1}\nabla_k^{(p-1)}\Phi^{(p)k} \label{equ:2a}\\
    \nabla^{(p)\mu}\nabla_{\mu}^{(p)} \Phi_{i}^{(p)}=\frac{1}{\sin^2{\theta_1}}\nabla^{(p-1)k}\nabla_k^{(p-1)} \Phi_{i}^{(p)}+(p-3)\cot{\theta_1}\partial_1\Phi_{i}^{(p)}+2\cot{\theta_1}\partial_i\Phi_{1}^{(p)} \nonumber\\
    \qquad\qquad\qquad+\frac{\Phi_{i}^{(p)}}{\sin^2{\theta_1}}-(p-1)\cot^2{\theta_1}\Phi_{i}^{(p)},~~(2\le i \le p) \label{equ:2b}
  \end{numcases}
\end{subequations}
\begin{subequations}
\label{equ:3}
  \begin{numcases}{}
    \nabla^{(p)\mu}\nabla_i^{(p)} \Phi_{\mu}^{(p)}=\partial_1(\nabla_k^{(p-1)} \Phi^{(p)k})+\partial_1^2 \Phi_1^{(p)}-(p-1)\cot^2{\theta_1} \Phi_1^{(p)} \nonumber\\
    \qquad\qquad\qquad+(p-1)\cot{\theta_1}\partial_1 \Phi_1^{(p)} \label{equ:3a}\\
    \nabla^{(p)\mu}\nabla_i^{(p)} \Phi_{\mu}^{(p)}=\nabla_k^{(p-1)}\nabla_i^{(p-1)} \Phi^{(p)k}+\partial_i\partial_1 \Phi_1^{(p)}+(p-1)\cot{\theta_1}\partial_i \Phi_1^{(p)} \nonumber\\
   \qquad\qquad\qquad +(1-(p-2)\cot^2{\theta_1})\Phi_i^{(p)},~~(2\le i \le p) \label{equ:3b}
  \end{numcases}
\end{subequations}
From \eq{equ:2a} and \eq{equ:3a}, we can easily get
\begin{equation}
  \label{equ:44}
  \nabla^{(p)\mu}(\nabla_{\mu}^{(p)} \Phi_{1}^{(p)}-\nabla_1^{(p)} \Phi_{\mu}^{(p)})=0
\end{equation}

Next, we  assume
\begin{equation}
  \label{equ:4}
  \nabla^{(p)\mu}(\nabla_{\mu}^{(p)} \Phi_{i}^{(p)}-\nabla_i^{(p)} \Phi_{\mu}^{(p)})=\mathcal{N}(L_p,p)\Phi_{i}^{(p)}.
\end{equation}
Substituting  \eq{equ:1}, \eq{equ:2b} and \eq{equ:3b} into \eq{equ:4}, we find
\begin{eqnarray}
  \label{equ:6}
  &&\nabla^{(p)\mu}(\nabla_{\mu}^{(p)} \Phi_{i}^{(p)}-\nabla_i^{(p)} \Phi_{\mu}^{(p)})  \nonumber\\
  &&=\displaystyle\sin^{\frac{4-p}{2}}{(\theta_1)}
   \left\{ \Phi_i^{(p-1)}\left[\partial_1^2+\cot{\theta_1}\partial_1
                               +\left(\frac{(p-2)(p-4)}{4}-\frac{1}{\sin^2{\theta_1}}(\frac{p-4}{2})^2\right)
                               \right] \right. \nonumber\\
  &&\left.+\frac{1}{\sin^2{\theta_1}}
            \left[\nabla^{(p-1)k}\left(\nabla_{k}^{(p-1)} \Phi_{i}^{(p-1)}
                                        -\nabla_i^{(p-1)} \Phi_{k}^{(p-1)}\right)
            \right]  \right\} \mathrm P_{L_p+\frac{p-2}{2}}^{L_{p-1}+\frac{p-2}{2}}(\cos{\theta_1}) \nonumber\\
  &&=\mathcal{N}(L_p,p)\sin^{\frac{4-p}{2}}{(\theta_1)} \mathrm P_{L_p+\frac{p-2}{2}}^{L_{p-1}+\frac{p-2}{2}}(\cos{\theta_1}) \Phi_i^{(p-1)}.
\end{eqnarray}
Since $\mathrm{P}_a^b$ is the associated Legendre function, satisfying
\begin{equation}
  \label{equ:7}
  \left[\partial_1^2+\cot{\theta_1}\partial_1+(a)(a+1)-\frac{ b^2}{\sin^2{\theta_1}}\right]\mathrm P_{a}^{b}(\cos{\theta_1})=0,
\end{equation}
the \eq{equ:6} reduces into
\begin{eqnarray}
  &&\mathcal{N}(L_p,p)=-(L_p+1)(L_p+p-2) \\
  &&\nabla^{(p-1)k}(\nabla_{k}^{(p-1)} \Phi_{i}^{(p-1)}-\nabla_i^{(p-1)} \Phi_{k}^{(p-1)})=-(L_{p-1}+1)(L_{p-1}+p-3)\Phi_i^{(p-1)} \label{equ:9}
\end{eqnarray}

In fact, the \eq{equ:44} can be considered as a special case of  \eq{equ:4}, and  \eq{equ:9} has the similar form as \eq{equ:4}. Remember that we do not assume $\Phi^{(p-1)}_i\ne 0$, one can repeat the same procedure to show that if  $\Phi_{\mu}^{(p)}$ takes the form
\begin{subequations}
  \begin{numcases}{}
    \Phi_i^{(p)}=0 ,~~(1\le i \le s-2)\\
    \Phi_i^{(p)}=\sin^{\frac{4-p}{2}}{(\theta_1)}\mathrm P_{L_p+\frac{p-2}{2}}^{L_{p-1}+\frac{p-2}{2}}(\cos{\theta_1})\Phi_i^{(p-1)},~~(s-2< i \le p) \\
    ...\\
    \Phi_i^{(j)}=\sin^{\frac{4-j}{2}}(\theta_{p-j+1})\mathrm P_{L_j+\frac{j-2}{2}}^{L_{j-1}+\frac{j-2}{2}}(\cos\theta_{p-j+1})\Phi_i^{(j-1)},~~(p-s+3\le j\le p),
  \end{numcases}
\end{subequations}
 they will satisfy
\begin{subequations}
  \begin{numcases}{}
  \nabla^k(\nabla_k \Phi_i^{(s;lm)}-\nabla_i \Phi_k^{(s;lm)})=0 ~,~(1\leq i\leq s-2) \\
  \nabla^{(j)k}(\nabla_{k}^{(j)} \Phi_{i}^{(j)}-\nabla_i^{(j)} \Phi_{k}^{(j)})=-(L_{j}+1)(L_{j}+j-2)\Phi_i^{(j)}~,~ (s-2<i\leq p)\label{sdiff}
  \end{numcases}
\end{subequations}

Thirdly, we need to prove a  set of functions $\Phi_i^{(N)}=\widetilde{Y_i}^{(lm)}(N),~(N=p-s+2)$
\begin{subequations}
  \begin{numcases}{}
  \Phi_{p-N+1}^{(N)}=H(\theta) Y_{lm}(\theta_{p-N+2},...,\theta_p) ;\label{equ:phyi}\\
  \Phi_i^{(N)}=\frac{\sin^2(\theta_{p-N+1})}{l(l+N-2)}\partial_i Y_{lm}(\theta_{p-N+2},...,\theta_p) \nonumber\\
  \times(\partial_{\theta_{p-N+1}}+(N-1)\cot{(\theta_{p-N+1})}) H(\theta) ;~(p-N+1<i\leq p )\\
  H(\theta)=\sin{(\theta_{p-N+1})}^{-\frac{N}{2}}P_{L_N+\frac{N-2}{2}}^{l+\frac{N-2}{2}}(\cos{\theta_{p-N+1}})
  \end{numcases}
\end{subequations}
satisfying \eq{sdiff}.
Applying  the definition \eq{equ:phyi} into following equation ( replacing $p$ by $N$ in \eq{equ:2} and \eq{equ:3} )
\begin{eqnarray}
  \label{equ:y1}
  &&\nabla^{(N)k}(\nabla_{k}^{(N)} \Phi_{s-1}^{(N)}-\nabla_{s-1}^{(N)}
  \Phi_{k}^{(N)})=\frac{1}{\sin^2{\theta_{s-1}}}[\nabla^{(N-1)k}\nabla_{k}^{(N-1)}+(N-1)]\Phi_{s-1}^{(N)} \nonumber\\
  &&+[\partial_{s-1}^2+(N+1)\cot{\theta_{s-1}}\partial_{s-1}-2(N-1)]\Phi_{s-1}^{(N)}
\end{eqnarray}
 and using the relation for scalar spherical harmonic function
\begin{equation}
  \nabla^{(N-1)k}\nabla_{k}^{(N-1)}Y_{lm}^{(N-1)}=-l(l+N-2)Y_{lm}^{(N-1)},
\end{equation}
one can arrive at
\begin{equation}
  \nabla^{(N)k}(\nabla_{k}^{(N)} \Phi_{s-1}^{(N)}-\nabla_{s-1}^{(N)}
  \Phi_{k}^{(N)})=-(L_N+1)(L_N+N-2)\Phi_{s-1}^{(N)},
\end{equation}
which completes the proof of \eq{sdiff} with $i=s-1$. For other $\Phi_i^{(N)},~(s-1<i\le p)$, the proof is almost same and  neglected.

In addition, the \eq{equ:as} need to be satisfied the gauge condition $G=0$ .
Since we have $A_t=A_r=0$, the gauge condition becomes into
\begin{equation}
  \label{equ:gc}
  \nabla^iA_i=0.
\end{equation}
One can easily show this equation  valid by using the definition of $A_i^{(p)}$ and $\Phi_i^{(p)}$.

\section{Acknowledgments}
This work has been supported by the National Natural Science Foundation of China (NSFC)
under Grant Nos. 11275099 and 11347605.

\end{CJK}
\end{document}